\documentclass[onecolumn,12pt,aps,prd,nobibnotes,showpacs,showkeys,
superscriptaddress]{revtex4-2}
\usepackage{graphicx,graphics,color,epsfig}
\usepackage[colorlinks=true,linkcolor=blue,citecolor=blue,urlcolor=blue,
breaklinks=true]{hyperref}
\usepackage{amsmath,amsfonts,amssymb}
\usepackage{verbatim}
\usepackage{latexsym}
\usepackage{dcolumn}
\usepackage{bm}
\usepackage{natbib}
\usepackage[english,russian]{babel}
\usepackage[utf8]{inputenc}

\begin{document}

\title{\href{}{Some scales in neutrino physics}}
\author{V.~V. Khruschov}
\email{khruschov{\_}vv@nrcki.ru}
\affiliation{National Research Center ``Kurchatov Institute'',
Kurchatov~place~1, 123182~Moscow, Russia}

\begin{abstract}
The problem of an active neutrinos mass origin  is considered
at the phenomenological level. At first an assumption is made that neutrino
mass values depend on two contributions with their characteristic
scales, a next assumption consists in their dependence on three dominant scales.  The neutrino mass values as well as the values of 
neutrino mass observables $m_C$ and $m_{\beta}$  are estimated.
\end{abstract}



\maketitle

 The problems of  experimental measurement of neutrino masses and determination
 of their origin remain among important problems of neutrino physics. 
It is known that neutrinos are considered as massless 
particles in the framework of the 
Standard Model (SM) and this fact was in a accordance with the experimental 
data in the past. But recently the contradictions had been arised between 
on the one hand the atmospheric and solar neutrino data and on the other hand 
the theoretical calculations. These
contradictions have been removed after an acceptance of neutrino 
oscillations and taking into account  Mikheev-Smirnov-Wolfenstein resonances
in the processes of neutrino interactions  with  matter.

At present the oscillations of flavor neutrinos are 
 confirmed by solid evidence from experiments
with atmospheric, solar, reactor and accelerator neutrinos   
\cite{mstv, gms,nava,este}. 
The oscillations of flavor neutrinos can be became realized with the help of 
mixing of neutrinos with different masses. Measuring of differences of
mass squares $\Delta m_{ij}^2=m_i^2-m_j^2$ and neutrino mixing parameters  
is carried out
in the oscillation experiments. However the neutrino mass absolute values cannot 
be discovered in these experiments as well as  Majorana or Dirac nature 
of neutrinos.

Three types of  experiments are sensitive to the absolute mass scale of 
neutrinos, namely: beta decay experiments, neutrinoless double beta decay
experiments, and some cosmological and astrophysical experiments. In each
type of these experiments a specific neutrino mass observable is measured.
These observables are the mean neutrino mass $m_C$,  the kinematical      
neutrino  mass $m_{\beta}$, and the effective neutrino mass $m_{2\beta}$,
which will be given below.

Mixing of three types of light neutrinos is defined with the help of 
the Pontecorvo-Maki-Nakagava-Sakata matrix:
\begin{equation}
\psi^{\alpha} = U^{\alpha}_i\psi^i,
\end{equation}
\noindent where $\psi^{\alpha,i}$ are left chiral fields of flavor or
massive neutrinos, $\alpha=\{e,\mu,\tau\}$, $i=\{1,2,3\}$.
$U$ is the Pontecorvo-Maki-Nakagava-Sakata matrix $U_{PMNS}=VP$, V can 
be written in the standard parametrization \cite{nava}
\begin{equation}
V = \left(\begin{array}{ccc}
  c_{12}c_{13} &     s_{12}c_{13} & s_{13}e^{-i\delta}\\
    -s_{12}c_{23}-c_{12}s_{23}s_{13}e^{i\delta}     &
c_{12}c_{23}-s_{12}s_{23}s_{13}e^{i\delta} & s_{23}c_{13}\\                
            s_{12}s_{23}-c_{12}c_{23}s_{13}e^{i\delta}  &
-c_{12}s_{23}-s_{12}c_{23}s_{13}e^{i\delta}  & c_{23}c_{13} 
\end{array}\right),
\end{equation}      
where  $c_{ij}  \equiv  cos\theta_{ij},  s_{ij} \equiv sin\theta_{ij}$,  
$P= \{\exp{i\alpha_{CP}}, \exp{i\beta_{CP}}, 1\}$, $\delta\equiv\delta_{CP}$ is the phase
connected with the Dirac CP non-conservation in the leptonic sector, while
$\alpha_{CP}, \beta_{CP}$ are the phases connected with the Majorana CP non-conservation.

The currently available oscillation parameters \cite{este} on the $1\sigma$ level,
 which determine  flavor oscillations  of three light neutrinos, are as follows
\[\sin^2{\theta_{12}}= 0.307^{+0.012}_{-0.011}\,,\]
\[\sin^2{\theta_{23}}=\left\{NO:0.561^{+0.012}_{-0.015}\atop IO:0.562^{+0.012}_{-0.015} \right.,\]
\[\sin^2{\theta_{13}}=\left\{NO:0.02195^{+0.00054}_{-0.00058}\atop 
IO:0.02224^{+0.00056}_{-0.00057} \right.,\]
\[\delta/^{\circ}=\left\{NO:177^{+19}_{-20}\atop 
IO:285^{+25}_{-28} \right.,\]
\[\Delta m_{21}^2/10^{-5}eV^2=7.49^{+0.19}_{-0.19}\,,\]
\begin{equation}
\Delta m_{31}^2/10^{-3}eV^2=\left\{NO:2.534^{+0.025}_{-0.023}\atop 
IO:-2.510^{+0.024}_{-0.025} \right..
\label{dat}
\end{equation}

\noindent The phases $\alpha_{CP}$ and $\beta_{CP}$ are not available
from the data at the moment as well as the absolute scale of neutrino
masses, for example, the $m_1$ mass magnitude.
 
As there is only the $\Delta m_{31}^2$ absolute value, then
the neutrino mass values can be arranged in two ways
\begin{equation}
\label{nh}
m_1<m_2<m_3,
\end{equation}
\begin{equation}
\label{ih}
m_3<m_1<m_2.
\end{equation}
\noindent The first case  of the arrangement of the neutrino masses (\ref{nh})
is named by the normal ordering of the neutrino mass spectrum (NO),
while the second one (\ref{ih}) is named by the inverted ordering (IO).

It is necessary to determine experimentally at the least one value among
neutrino mass observables $m_C$,   $m_{\beta}$ or $m_{2\beta}$,
in order to know the  active neutrino absolute mass scale.
\begin{equation}
\label{mc}
3\,m_C= \sum_{i=1,2,3}|m_i|,
\end{equation}
\begin{equation}
\label{mb}
m_{\beta}^2=\sum_{i=1,2,3}|U_{ei}|^2m_i^2,
\end{equation}
\begin{equation}
\label{mbb}
m_{2\beta}=\left|\sum_{i=1,2,3}U_{ei}^2m_i\right|.
\end{equation}

Effective neutrino mass $m_{2\beta}$ is the upper diagonal
matrix element of the mass matrix for Majorana neutrinos
\begin{equation}
M=Um^dU^T,
\end{equation}
where $m^d=diag\{m_1,m_2,m_3\}$. Absolute values of two additional diagonal matrix elements 
$m_{\mu\mu}$ and $m_{\tau\tau}$ are  most likely  equal each other. 
This supposition does not contradict to 
 some models for the neutrino mass matrix
\cite{fri, jo}.

The following experimental limits for  neutrino mass observables are
obtained at 90\% CL \cite{este}: $m_C<0.013-0.1eV$ \cite{ji,nar,desi}, $m_{\beta}<0.45 eV$ \cite{katr}, 
$m_{2\beta}<0.028-0.122 eV$ \cite{gerd,kaml},
where the last limit should be increased up to $0.079-0.180$ eV in order to include
the uncertainty of nuclear matrix element values. Note that the listed limits on the mass observables values (\ref{mc}), (\ref{mb}), (\ref{mbb}) are not in contradiction with the given previously neutrino mass values and the NO of their spectrum \cite{14b, yud}, namely $m_1 \approx 0.0016$ eV, $m_1 \approx 0.0088$ eV, 
$m_1 \approx 0.0496$ eV. In addition the results of the latest cosmological observations \cite{desi,nar,37b,37c,37d,37e} justify the NO option with the total sum of the active neutrinos masses about 0.06 eV.

The important question is to find out a mechanism of generation of
neutrino masses. Lacking a satisfactory theory of this phenomenon,
the question can be treated on the phenomenological level. At first let us suppose
that there are a few different contributions in a neutrino mass, and 
two of them are most important. It may be assumed that the first 
contribution is connected with the Majorana mass of the light left 
neutrino beyond the SM. This contribution  can arise
on  some characteristic scale due to the presence in a lagrangian an effective Majorana mass term
when the Higgs sector of SM is modified.
\begin{equation}
L'_m=-\frac{1}{2}\overline{\nu_{L}}M_{\nu}\nu^c_{L} + h.c.
\end{equation}

The contribution in a neutrino mass value, that is connected with $L'_m$,
will be taken into account through the phenomenological parameter $\xi$.
The second contribution can be connected with the {\it seesaw} mechanism,
that takes place when one adds heavy right (sterile) neutrinos 
$N_i, i=1,2,3$ in this scheme. This mass contribution can be written
 in the form:
\begin{equation}
M''_{\nu}=-M_D^TM_R^{-1}M_D,
\end{equation}
\noindent where $M_D$ is the matrix of Dirac neutrino masses, $M_R$
is the characteristic mass of the right neutrinos.

Thus, the new scale connected with the masses of heavy sterile neutrinos
 appears. Let us assume that $M_D$ is proportional to the mass matrix
for charged leptons:
\begin{equation}
M_D = \sigma M_l,
\end{equation}
\noindent where a $\sigma$ magnitude is of the order of unity, 
$M_l=diag\{m_e, m_{\mu}, m_{\tau}\}$, $M_R\sim M$. 
So the following phenomenological
formula will be used for estimations of neutrino masses:
\begin{equation}
m_{\nu i}=\pm\xi-\frac{m_{li}^2}{M}.
\label{for}
\end{equation}

Taking into account the data (\ref{dat}) the absolute values of
neutrino masses  $\mu_i$ and characteristic scales $\xi$ and $M$ in $eV$
have been obtained in the NO case.
\begin{equation}
NO: \mu_1\approx0.0693, \; \mu_2\approx0.0698, \; \mu_3\approx0.0851, \;
\xi\approx0.0693, \; M\approx2.0454\times10^{19}.
\label{mnh}
\end{equation}

However one can  simplify the mass formulae (\ref{for}). If the structure of
the Standard Model Higgs sector is unchanged, then a $\xi$ value may be neglected.
Using three mass values of heavy sterile neutrinos $M_1$, $M_2$, $M_3$
and $\sigma=i/\sqrt{2}$, we obtain 
\begin{equation}
\label{forn}
m_{\nu i}=\frac{m_{li}^2}{2M_i}.
\end{equation}
\noindent When the neutrino mass values written above are substituted in 
Eq.~(\ref{forn}) the following values are obtained  
\begin{equation}
M_1\approx81.6 \; TeV, M_2\approx0.6343 \; EeV, M_3\approx31.8294 \; EeV.
\label{mne}
\end{equation}

Now we can evaluate the neutrino mass observables $m_C$ and $m_{\beta}$ in $eV$
($m_{2\beta}$ depends on unknown  $\alpha_{CP}$ and $\beta_{CP}$ phases).
 \begin{equation}
 \label{enh}
NO: m_C \approx 0.02, \quad m_{\beta}\approx 0.01,
\end{equation}
\noindent which do not contradict the written above limits on these mass
observables.

In summary it is suggested the simple formula (\ref{forn}) for
evaluations of neutrino masses which depends on three characteristic
scales $M_1$, $M_2$ and $M_3$ (\ref{mne}). Calculated 
$m_C$ and $m_{\beta}$ evaluations   (\ref{enh})
can be used for predictions and interpretations of neutrino data.

\smallskip

\begin{center}
\textbf{Acknowledgements}
\end{center}

This work was supported by the RFBR under grant 11-02-00882-a.

\end{document}